\documentclass[showpacs,twocolumn,pra]{revtex4}%
\usepackage{amssymb}
\usepackage{bm}
\usepackage{amsmath}
\usepackage{graphicx}
\usepackage{amsfonts}%
\begin{document}
\title{Ultra-high-Q toroidal microresonators for cavity quantum electrodynamics}
\author{S. M. Spillane, T. J. Kippenberg, K. J. Vahala}
\affiliation{Thomas J. Watson Laboratory of Applied Physics, California Institute of
Technology, Pasadena, CA 91125}
\author{K. W. Goh, E. Wilcut, H. J. Kimble}
\affiliation{Norman Bridge Laboratory of Physics, California Institute of Technology,
Pasadena, CA 91125}
\date{\today}

\begin{abstract}
We investigate the suitability of toroidal microcavities for strong-coupling
cavity quantum electrodynamics (QED). Numerical modeling of the optical modes
demonstrate a significant reduction of modal volume with respect to the
whispering gallery modes of dielectric spheres, while retaining the high
quality factors representative of spherical cavities. The extra degree of
freedom of toroid microcavities can be used to achieve improved cavity QED
characteristics. Numerical results for atom-cavity coupling strength, critical
atom number $N_{0}$ and critical photon number $n_{0}$ for cesium are
calculated and shown to exceed values currently possible using Fabry-Perot
cavities. Modeling predicts coupling rates $g/2\pi$ exceeding 700 MHz and
critical atom numbers approaching $10^{-7}$ in optimized structures.
Furthermore, preliminary experimental measurements of toroidal cavities at a
wavelength of 852 nm indicate that quality factors in excess of 100 million
can be obtained in a 50 micron principal diameter cavity, which would result
in strong coupling values of $(g/(2\pi),n_{0},N_{0})=(86$ MHz$,4.6\times
10^{-4},1.0\times10^{-3})$.

\end{abstract}
\maketitle

\section{Introduction}

The use of an optical microcavity can greatly enhance the interaction of an
atom with the electromagnetic field such that even a single atom or photon can
significantly change the dynamical evolution of the atom-cavity system
\cite{Kimble-cQED}. Achieving the regime of "strong coupling"
\cite{KimbleCQED_book,Mabuchi}\ is critically dependent on the characteristics
of the optical cavity and generally requires the optical modes to be confined
in a small mode volume for extended periods of time (or equivalently high Q-factor).
 
Recent experimental realizations of strong coupling have employed
high-finesse Fabry-Perot (FP) optical microcavities \cite{Hood00,Pinkse00,Shimizu02,PRLTrapping,Maunz04,Sauer04}. Our experiments at Caltech include the realization of an "atom-cavity microscope" with a single atom bound in orbit by single photons \cite{Hood00}, and the development of a laser that operates with
"one and the same" atom \cite{SingleAtomLaserKimble}. Fabry-Perot cavities, while
possessing ultra-high quality factors and finesse, are difficult to
manufacture and control, requiring sophisticated dielectric mirror coatings as
well as accurate feedback for resonant wavelength control. Due in part to
these reasons, there has been increased interest in other microcavity systems
which not only can address some or all of the limitations of Fabry-Perot
cavities, but which in principle can have improved optical properties.

Based upon the pioneering work of V. Braginsky and colleagues
\cite{Nonlinear_spheres_Ilchenko}, whispering-gallery-mode cavities have also
been investigated for cQED experiments for many years \cite{Vahala03}.
Experimental studies have demonstrated Q-factors approaching $10^{10}$ in a
silica microsphere whispering gallery cavity \cite{Gorodetsky96,Vernooy}, with
values exceeding $10^{8}$ readily achievable over a broad range of cavity
diameters and wavelengths. The combination of their very low cavity losses,
small mode volumes and their relative ease of fabrication, makes them
promising candidates for experiments in cQED \cite{Haroche_CQED,Vernooy_CQED}.
Furthermore, the ability to couple these cavities with record coupling
efficiencies to an optical fiber \cite{Ideality} (the medium of choice for low
loss transport of classical and nonclassical states \cite{Cirac}) is
fundamentally important in cQED and bears promise for realizing quantum networks.

Recently, a new type of whispering-gallery-mode optical microcavity was
demonstrated, which not only retains the high quality factors of spherical
cavities, but also has significant advantages in fabrication reproducibility,
control, and mode structure. These cavities consist of a toroidally-shaped
silica cavity supported by a silicon pillar on a microelectronic chip
\cite{Toroid}. The toroidal cavity shape allows an extra level of geometric
control over\ that provided by a spherical cavity and thus begs the question
on how these structures compare with silica microspheres and other microcavity
designs for strong-coupling cavity QED. In this manuscript we numerically
investigate the suitability of toroidal microcavities for strong-coupling
cavity QED experiments, and, for purposes of comparison, we focus on the
interaction with atomic cesium \cite{Hood00,FP}. We show that toroid
microcavities can achieve ultra-high-quality factors exceeding 100 million
while simultaneously obtaining very large coupling rates between the cavity
and a cesium atom. It is found that these cavities not only surpass the
projected limits of FP technology \cite{FP}, but also either exceed or compare
favorably to other cavity designs such as photonic bandgap devices
\cite{PBGJelenaCQED,PBGPainterMabuchi}. Lastly, we present preliminary
experimental measurements of quality factors for toroidal cavities at a
wavelength of 852 nm, suitable for strong-coupling cQED with atomic cesium.
These results show that currently attainable Q values are already quite promising.

\section{Strong-coupling in an atom-cavity system}

The coupling rate $g$\ between an atomic system and an electromagnetic field
is related to the single-photon Rabi frequency $\Omega=2g$, and can be
expressed in terms of the atomic and cavity parameters by \cite{Kimble-cQED},%
\begin{align}
g(\mathbf{r})  &  =\gamma_{\perp}|\vec{E}(\mathbf{r})/\vec{E}_{max}%
|\sqrt{V_{a}/V_{m}}\label{g_equation}\\
V_{a}  &  =3c\lambda^{2}/(4\pi\gamma_{\perp})
\end{align}
where $\gamma_{\perp}$ is the transverse atomic dipole transition rate,
$|\vec{E}(\mathbf{r})/\vec{E}_{max}|$ denotes the normalized electric field
strength at the atom's location $\mathbf{r}$, $V_{a}$ is a characteristic
atomic interaction volume (which depends on the atomic dipole transition rate,
the transition wavelength $\lambda$, and the speed of light $c$), and $V_{m}$
is the cavity electromagnetic mode volume. Assuming the atom interacts with
the electromagnetic field for a time $T$, strong atom-field coupling occurs if
the rate of coupling exceeds all dissipative mechanisms, i.e. $g\gg
(\kappa,\gamma_{\perp},T^{-1})$. In this expression $\kappa$ denotes the
cavity field decay rate, given in terms of the cavity quality factor $Q$ by
$\kappa\equiv\pi c/(\lambda Q)$. The degree of strong coupling can also be
related to a set of normalized parameters \cite{Kimble-cQED},%
\begin{align}
n_{0}  &  \equiv\gamma_{\perp}^{2}/(2g^{2})\\
N_{0}  &  \equiv2\gamma_{\perp}\kappa/(g^{2})
\end{align}
where $n_{0}$ is the critical photon number, which is the number of photons
required to saturate an intra-cavity atom, and $N_{0}$ is the critical atom
number, which gives the number of atoms required to have an appreciable effect
on the cavity transmission. Note that $(N_{0},n_{0})\ll1$ provides a necessary
but not sufficient condition for strong coupling.

Examining these parameters, we see that only the critical atom number
$N_{0}\propto V_{m}/Q$\ is dependent on the cavity loss rate (or equivalently
Q factor). It is the possibility of realizing extremely low critical atom
numbers with ultra-high-Q microcavities that has fostered the investigation of
silica microspheres for strong-coupling cQED experiments. However, the
geometry of a spherical dielectric dictates a definite relationship between
cavity mode volume $V_{m}$\ and the associated quality factor $Q$, and hence
of the value of the coupling parameter $g\propto V_{m}^{-1/2}$ while still
maintaining ultra-high quality factors \cite{BuckPRA}. This is a result of the
fact that to achieve large atom-cavity coupling rates (comparable to or
exceeding those of FP\ cavities) the cavity diameter must be made small (8
micron diameter sphere gives $g/(2\pi)\approx740$ MHz) in order to both lower
the modal volume and to increase the electric field strength at the atomic
position (assumed to be the cavity surface at the point of maximum electric
field strength). However, at the optimum radius for atom-coupling strength,
the tunneling loss of the microcavity results in a low achievable Q-factor
($Q\approx4\times10^{4}$), thereby raising the critical atom number. While the
relatively large mode volumes of silica microsphere cavities preclude them
from competing with ultra-small mode volume cavities (such as photonic bandgap
cavities) on the basis of coupling strength alone, there is the possibility to
access simultaneously both ultra-high-Q and small mode volume, using toroidal microresonators.

\section{Toroidal microresonators}

\begin{figure}[ptb]
\begin{center}
\includegraphics[scale=0.75]{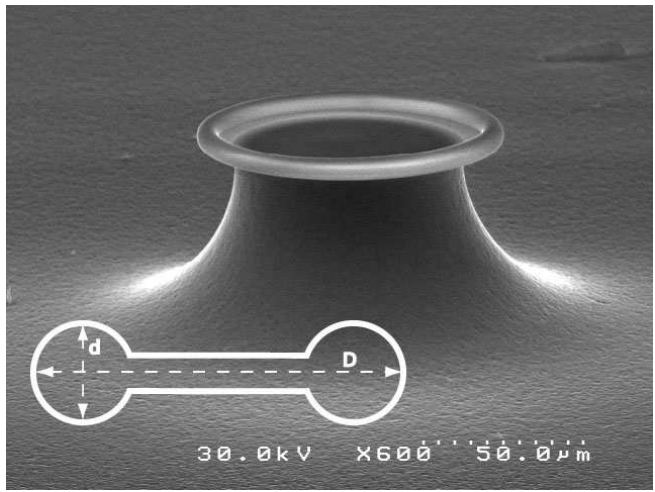}
\end{center}
\caption{Scanning electron micrograph of a toroidal microcavity. The principal
and minor diameters are denoted by $D$ and $d$, respectively.}%
\label{SEM}%
\end{figure}

\begin{figure*}[ptb]
\begin{center}
\includegraphics[scale=1]{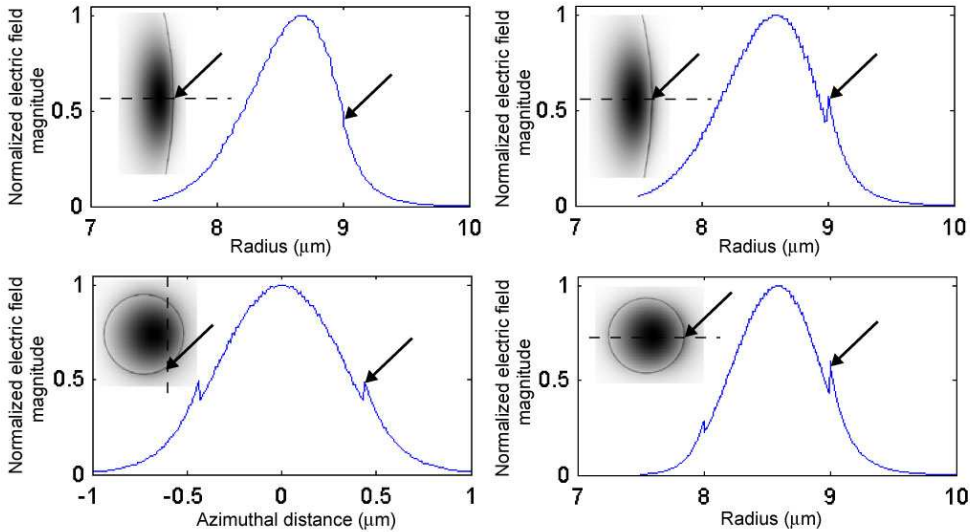}
\end{center}
\caption{Electric field magnitude for the whispering-gallery modes of a
spherical (top row) cavity of diameter 18 microns and a toroidal cavity
(bottom row) with principal diameter of 18 microns and a minor diameter of 1
micron. The left (right) column shows the TE (TM) polarized mode near 850 nm.
The arrows indicate the location of the maximum external electric field
strength, where we assume the atom is located. The dotted lines in the
two-dimension field distribution indicate the cross-section where the electric
field is displayed.}%
\label{TE_TM_panel}%
\end{figure*}

Toroidal microresonators are chip-based microcavities that possess
ultra-high-Q ($>$100 million) whispering-gallery type modes\cite{Toroid}. The
realization of ultra-high-Q chip-based resonators allows improvements in
fabrication and control, while additionally allowing integration with
complementary optical, mechanical or electrical components. In brief, these
resonators are fabricated by standard lithographic and etching techniques,
followed by a laser-reflow process, as outlined in reference \cite{Toroid}.
The combination of thermal isolation of the initial preform periphery and
thermal heat sinking of the preform interior through the strong heat
conduction of the silicon support pillar results in a preferential melting of
the preform along the disk periphery under $CO_{2}$ laser irradiation. Surface
tension then induces a collapse of the silica disk preform, resulting in a
toroidally-shaped boundary, with the final geometry controlled by a
combination of irradiation flux and exposure time. Importantly, as the optical
mode resides in the extremely uniform and smooth (reflowed) periphery of the
structure, the quality factors of optical whispering-gallery modes can achieve
ultra-high-Q performance, exceeding 100 million. Figure \ref{SEM} shows a
scanning electron micrograph of the side-view of a typical toroidal
microcavity. Quality factors as high as 400 million at a wavelength of 1550 nm
(corresponding to a photon lifetime of $\sim$300 ns) have been measured
\cite{UHQ_toroid_APL}.

\section{Microtoroid numerical modeling}

In order to investigate the properties of microtoroids for cQED, this paper
will focus on the $D_{2}$ transition of cesium which occurs at a wavelength of
852.359 nm \cite{FP}, with scaling to other systems accomplished in the
fashion of Ref. \cite{BuckPRA}. Fundamentally, the coupling between an atom
and a cavity field can be specified by three parameters: the cavity field
strength at the atom's location, the cavity mode volume $V_{m}$, and the
cavity quality factor $Q$. Since the optical modes are confined to the
interior dielectric in whispering-gallery-type resonators, the atom can
interact only with the evanescent field of the cavity mode. In the following
discussion, the atom is assumed to be located near the resonator surface at
the location where the electric field strength is largest, as illustrated in
figure \ref{TE_TM_panel}. For TM\ polarized modes (defined such that the
dominant electric field component is in the radial direction) this occurs at
the outer cavity boundary in the equatorial plane, while for TE polarized
modes (dominant electric field component in the azimuthal/vertical direction)
the location of the maximum external field strength is more complicated. As
the toroidal geometry is compressed with respect to a sphere (i.e. reducing
the ratio of minor-to-principal toroid diameter), the maximum field strength
for a TE polarized mode changes from the equatorial outer cavity boundary to
approaching the azimuthal axis (see figure \ref{TE_TM_panel}). While the
precise localization of the atom at the cavity evanescent field maximum has
been analyzed in detail \cite{Mabuchi94,VernooyAtomGallery}, such localization
has not yet been achieved experimentally. Nonetheless, this assumption allows
a simple way to characterize the relative merit of this cavity geometry with
respect to other cavity designs. Also, in what follows we will only consider
the fundamental radial and azimuthal modes for both polarizations (TE and TM),
as they possess the smallest modal volumes and thus the highest coupling strengths.

\begin{figure}[ptb]
\begin{center}
\includegraphics[scale=1.4]{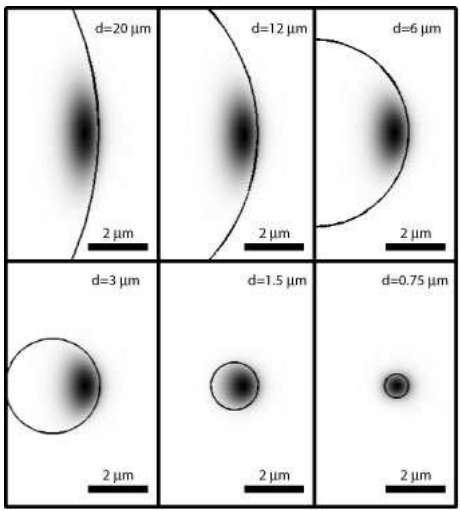}
\end{center}
\caption{Electric field profiles for a toroidal cavity with a principal
diameter $D=$ 20 microns and minor diameters $d=$ 20, 12, 6, 3, 1.5, and 0.75
microns. The calculations correspond to a TM polarized mode near 850 nm. The
optical mode behaves as a whispering-gallery-type mode until the minor
diameter is below approximately 1.5 microns, at which point the mode
approaches that of a step-index optical fiber \cite{Optical_Waveguide_Theory}%
.}%
\label{mode_plots}%
\end{figure}

The microtoroid geometry, which exhibits a dumbbell-shaped cross-section, can
in most cases be considered a torus, as the presence of the supporting disk
structure only affects the optical mode when the torus diameter becomes
comparable to the radial extent of the optical mode. As shown in figure
\ref{mode_plots}, this point occurs when the toroid minor diameter (i.e. the
cross-sectional diameter of the torus) is below approximately 1.5 microns for
a principal diameter of 16 microns. Furthermore, through improvements in
fabrication the influence of the toroid support can in principle be minimized.
In contrast to FP and microsphere cavities, the optical modes of a toroid do
not possess analytic solutions. While one can derive approximate expressions
for the optical behavior of these structures for both the low transverse
compression (sphere-like) and high transverse compression (step-index,
fiber-like) regimes, we are mostly interested in the intermediate geometrical
regimes, as these are both experimentally accessible and retain the most
desirable properties of whispering-gallery-type microcavities. To accomplish
this task, a two-dimensional finite element eigenmode/eigenvalue solver was
used to characterize the optical modes of the cavity over the complete
geometrical range, after explicitly accounting for the rotational symmetry.
The optical modes were calculated in a full-vectorial model, which provides
the complete electric field dependence. The accuracy of the numerical
technique was carefully verified by comparison with results using the
analytical solution for a microsphere cavity \cite{Stratton}. The results for
the mode volumes, resonance wavelengths, and field profiles were in good
agreement (fractional error was less than $10^{-4}$ and $10^{-2}$ for the
resonance wavelength and modal volume, respectively). Furthermore, the error
in the radiation quality factor was less than 10\% over a wide value of
radiation Q's ($10^{3}$ to $10^{14}$), demonstrating that this method can give
the accuracy required to investigate the fundamental radiation loss limits in
the cavity geometries of interest in this work. Due to the fact that for
smaller cavity geometries the resonance wavelengths do not necessarily
coincide with the cesium transition of interest, the data in this work were
evaluated by using values calculated at the closest resonance wavelengths,
both blue and red-shifted with respect to the desired resonance, to
extrapolate values at the desired wavelength (the mode volumes were linearly
extrapolated and the radiation quality factors exponentially extrapolated as a
function of wavelength).

\subsection{Mode volume}

The optical mode volume is determined by,
\begin{equation}
V_{m}\equiv\frac{\int_{V_{Q}}\epsilon(\vec{r})|\vec{E}(\vec{r})|^{2}d^{3}%
\vec{r}}{|\vec{E}_{\max}|^{2}}%
\end{equation}
where $V_{Q}$ represents a quantization volume of the electromagnetic field,
and $|\vec{E}|$ is the electric field strength \cite{VernooyAtomGallery}. In
these calculations, we have chosen the quantization volume cross-section to
consist of a square region of approximately 10 micron width and height
centered about the radial cavity boundary. This choice allows the mode volume
to be determined to a good accuracy while minimizing computational
requirements. As a further confirmation of the validity of this approach, we
note that the radiation loss is weak for the range of geometries modeled in
this work, resulting in only a marginal difference in the numerically
calculated mode volume for different choices of quantization volume.

\begin{figure}[ptb]
\begin{center}
\includegraphics[scale=0.75]{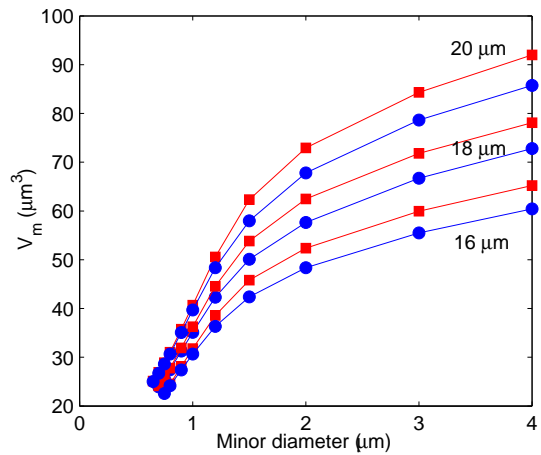}
\end{center}
\caption{Calculated mode volumes for a silica toroidal microresonator versus
minor diameter for principal diameters of 20, 18, and 16 microns. The plot
shows both TM (red squares) and TE (blue circles) polarizations. As the minor
diameter is reduced a slow reduction of modal volume due to confinement in the
azimuthal direction occurs, followed by a fast reduction for large confinement
when the optical mode is strongly compressed in both the radial and azimuthal
directions.}%
\label{mode_volume}%
\end{figure}

Figure \ref{mode_volume} shows the calculated modal volume for the fundamental
mode of a toroidal cavity as a function of minor diameter and for principal
diameters ranging from 16 to 20 microns. For clarity, only data for minor
diameters below 4 microns are shown. Both TM (squares) and TE (circles)
polarizations are shown. The calculations show a reduction of modal volume for
both polarizations as the toroid minor diameter is decreased. This is expected
when considering the additional confinement provided by the toroid geometry
beyond the spherical geometry, as illustrated in the electric field plots of
figure \ref{mode_plots}. As the minor diameter is decreased, there is
initially a slow reduction of modal volume, which agrees very well with a
simple model that accounts for transverse guiding (azimuthal direction) using
an approximate one-dimensional harmonic oscillator model. This approach
results in a reduction of modal volume which scales as $(d/D)^{1/4}$ with
respect to that of a spherical cavity. This formula holds for minor diameters
greater than approximately 2 microns for the principal diameters considered in
this work. For smaller diameters, the spatial confinement becomes strong
enough that the optical mode is additionally compressed in the radial
direction. This results in a faster reduction of modal volume, with the
optical modes approaching those of a step-index optical fiber (this occurs for
a minor diameter below approximately 1 micron) \cite{Optical_Waveguide_Theory}%
. The mode volume reduces until the point where the optical mode becomes
delocalized due to the weak geometrical confinement, causing a finite minimum
value. Determination of the exact point of the minimum modal volume upon
reduction of minor diameter (for a fixed principal diameter) can be uncertain,
as the choice of quantization volume now plays a critical role (as discussed
above). For this reason the results in figure \ref{mode_volume} show the modal
volume only for inner diameters down to 0.65 microns, where mode volume
determination was unambiguous.

Calculation of the modal volume and the maximum electric field amplitude at
the exterior cavity equatorial boundary is straightforward, giving a simple
way to calculate both the coupling strength and the critical photon number. In
order to obtain the cavity decay rate $\kappa$ and the critical atom number
$N_{0}$, however, the cavity Q factor must be determined.

\subsection{Quality factor}

The radiation loss of the optical modes of a spherical cavity is easily found
by consideration of the analytic characteristic equation \cite{Weinstein},
\begin{equation}
n^{1-2b}\frac{[nkRj_{\ell}(nkR)]^{\prime}}{nkRj_{\ell}(nkR)}=\frac{[kRh_{\ell
}^{(1)}(kR)]^{\prime}}{kRh_{\ell}^{(1)}(kR)}%
\end{equation}
where $n$ is the refractive index of the spherical cavity (the external index
is assumed to be unity), $R$ is the cavity radius, $b$ represents the
polarization of the optical mode (1 for TM and 0 for TE), and $j_{\ell}$
$(h_{\ell}^{(1)})$ represent the spherical Bessel (Hankel) functions. The
prime denotes differentiation with respect to the argument of the Bessel
(Hankel) functions. This equation accounts for radiation loss through the use
of an outgoing wave outside the cavity, as given by the complex Hankel
function of the first kind. Solution of this equation results in a complex
wavenumber, $k=k_{Re}+ik_{Im}$, which determines both the resonance wavelength
($\lambda=2\pi/k_{Re}$) and the radiation quality factor ($Q_{rad}%
=k_{Re}/(2k_{Im})$).

\begin{figure}[ptb]
\begin{center}
\includegraphics[scale=0.75]{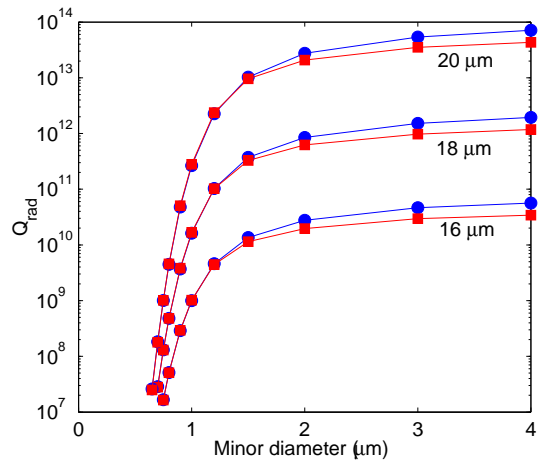}
\end{center}
\caption{Calculated radiation loss for a toroidal microcavity as a function of
minor diameter, for principal diameters of 20, 18, and 16 microns. Both TM
(red squares) and TE (blue circles) polarizations are shown. The data show a
slow reduction of Q as the minor diameter is reduced while the mode behaves
primarily as a whispering-gallery-type mode. However, as the geometrical
confinement increases to such a point as the optical mode approaches that of a
step-index fiber, there is a significant reduction of quality factor.}%
\label{Q_radiation}%
\end{figure}

However, while the spherical solution can provide some insight on the scaling
of the radiation quality factor for toroidal cavities where the minor diameter
is large (sphere-like), the radiation loss when the optical mode is strongly
confined (as represented by small minor diameters) is expected to decrease
much more rapidly. Figure \ref{Q_radiation} shows numerical calculations of
the radiative quality factor as the minor diameter is decreased for various
principal diameters of 16, 18, and 20 microns. We observe an initially slow
reduction of the radiative quality factor in the geometrical regime where the
minor diameter exceeds the radial extent of the optical mode (i.e., where the
optical mode exhibits whispering-gallery behavior). As the minor diameter is
reduced to a level comparable to or smaller than the radial extent of the
optical mode (step-index fiber-like regime), the drop-off of the radiative Q
is much more dramatic, with a decrease of over an order of magnitude for a
reduction of inner diameter of just 50 nm.

The total optical loss of a cavity has contributions not only from radiation
loss, but also includes other dissipative mechanisms, such as intrinsic
material absorption, losses resulting from both surface and bulk scattering,
and losses stemming from contaminates on the resonator surface
\cite{GorodetskyUltimateQ}. One of the dominant contaminates which adversely
affects the cavity Q is OH\ and water adsorbed onto the cavity surface. While
prior investigations on these loss mechanisms have resulted in approximate
expressions for water absorption and surface scattering \cite{Vernooy,Gorodetsky_Rayleigh}, only very large resonators were studied, as
opposed to the much smaller diameter cavities studied in this work. To obtain
an improved estimate of the effect of water on the small diameter cavities in
this manuscript, a simple model was used which determines the fraction of
optical energy absorbed by a monolayer of water located at the cavity surface.
This method gives an estimated quality factor for a monolayer of water to be
greater than $10^{10}$ for the case of a spherical resonator with a principal
diameter of 50 microns. While the water-limited quality factor will be
slightly lower for the smaller principal diameter cavities in this work, and
also slightly lower due to the increased overlap between the optical mode and
the cavity surface in a toroidal geometry, these values are comparable to the
quality factor due solely to the intrinsic absorption of silica in the 800 nm
wavelength band. As in principle with proper fabrication the presence of water
and OH can be prevented, with surface scattering minimized, we will focus only
on the contributions from intrinsic silica absorption and radiation loss.
These two mechanisms put a fundamental limit on the Q possible in these structures.

\begin{figure}[ptb]
\begin{center}
\includegraphics[scale=0.75]{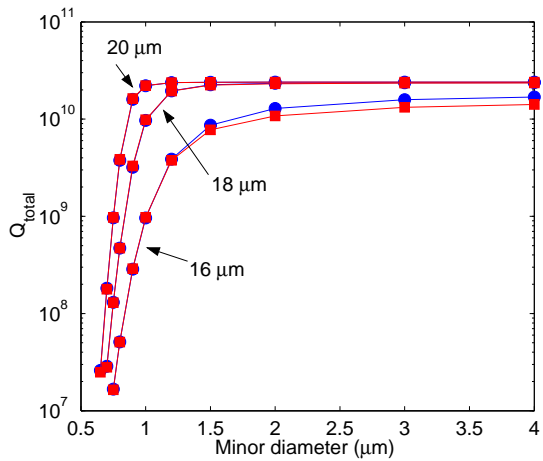}
\end{center}
\caption{Total quality factor for a toroidal microcavity versus minor diameter
for principal diameters of 16, 18, 20 microns. Both TE (blue circles) and TM (red
squares) polarizations are shown.  The total quality factor is
composed of the radiative quality factor from figure \ref{Q_radiation} along
with the silica absorption limited $Q_{mat}=2.4\times10^{10}$ at a wavelength
of 852 nm. The plots indicate that the total quality factor is limited by
silica absorption when the principal diameter is larger than 16 microns and
the minor diameter is larger than approximately 1 micron. Furthermore, both polarizations have similar quality factors over the range of geometries
studied.}%
\label{Q_total}%
\end{figure}

Figure \ref{Q_total} shows the calculated total quality factor for various
principal toroid diameters in the range of 16-20 microns, as a function of the
minor diameter. The total quality factor is calculated through the relation
$1/Q_{total}=1/Q_{rad}+1/Q_{mat}$, where only radiation loss and silica
absorption are included. For principal diameters less than 18 microns, there
is a monotonic decrease in quality factor as the minor diameter is decreased.
This is a result of the whispering-gallery-loss increase due to the additional
confinement. For larger principal diameters, the overall quality factor is
clamped near the limiting value resulting from silica absorption for most
minor diameters (with only a slight decrease as minor diameter is reduced),
until the minor diameter is small enough that the radiative quality factor
decreases below the quality factor due to silica absorption. For the principal
diameters studied in this work, this point occurs as a minor diameter of
around 1 micron.

\subsection{Cavity QED parameters}

\begin{figure}[ptb]
\begin{center}
\includegraphics[scale=0.85]{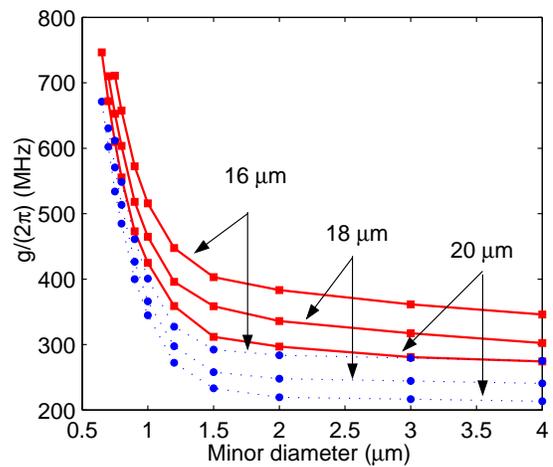}
\end{center}
\caption{Atom-cavity coupling parameter $g$ vs. minor diameter for toroidal
cavities having a principal diameter of 16, 18, and 20 microns, with $g$
increasing for smaller principal diameters. Both TE (blue circles) and TM (red
squares) polarizations are shown. The plots indicate that the coupling
strength increases dramatically as the minor diameter decreases below 1.5
microns, which is a result of the rapid reduction of mode volume and the
increased electric field strength at the cavity surface.}%
\label{g}%
\end{figure}

The determination of the coupling strength from the modal volume follows from
equation \ref{g_equation}. Figure \ref{g} shows the atom-cavity coupling rate
$g/(2\pi)$ for various toroid principal diameters as the toroid minor diameter
is decreased. It can be seen that there is a monotonic rise in $g$ for
higher-aspect ratio toroids (i.e. $D/d$), as a direct result of the
compression of modal volume. The rate of increase of $g$ as the minor diameter
is reduced increases dramatically as the toroid geometry transitions from a
whispering-gallery-type mode to a strongly-confined step-index fiber-type
mode. This is due not only to the faster rate of reduction of mode volume in
the step-index fiber-like regime as the minor diameter is decreased, but also
due to the increase in electric field strength at the cavity surface (as
$g\propto|E|(V_{m})^{-1/2}$). Note that the coupling strengths shown do not
correspond to the absolute maximum for these structures, as this work has
focused on the simultaneous realization of high quality factors and small
modal volume. Therefore, mode volumes were calculated only down to where the
radiation quality factor is equal to or slightly exceeds 10 million. Also, as
mentioned previously, by making this restriction we prevent any uncertainty in
the calculated mode volumes (and hence $g$) through the definition of the
modal quantization volume. Under these assumptions, the calculations indicate
that coupling parameters exceeding 700 MHz are possible.

\begin{figure}[ptb]
\begin{center}
\includegraphics[scale=0.85]{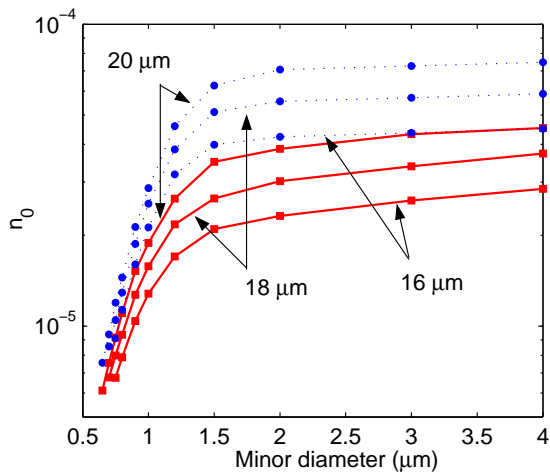}
\end{center}
\caption{Critical photon number $n_{0}$\ vs. minor toroid diameter for a
cavity with principal diameters of 16, 18, and 20 microns. Both TE (blue
circles) and TM (red squares) polarizations are shown. The plots show that as
both toroid principal diameter and minor diameter are reduced, the critical
photon number decreases. This follows directly from the behavior of the
atom-cavity coupling parameter $g$, as indicated in figure \ref{g}. The
calculations show that critical photon numbers of $6\times10^{-6}$ are
possible (with quality factors exceeding 10 million).}%
\label{critical_photon}%
\end{figure}

Figure \ref{critical_photon} shows the corresponding critical photon numbers
($n_{0}$). The results reveal that values as low as $6\times10^{-6}$ are
possible, with the associated quality factors exceeding 10 million. As will be
discussed in more detail in the next section, this value is not only
comparable to the fundamental limit of FP technology, but also vastly exceeds
that possible for fused silica microspheres with a comparable quality factor.

\begin{figure}[ptb]
\begin{center}
\includegraphics[scale=0.85]{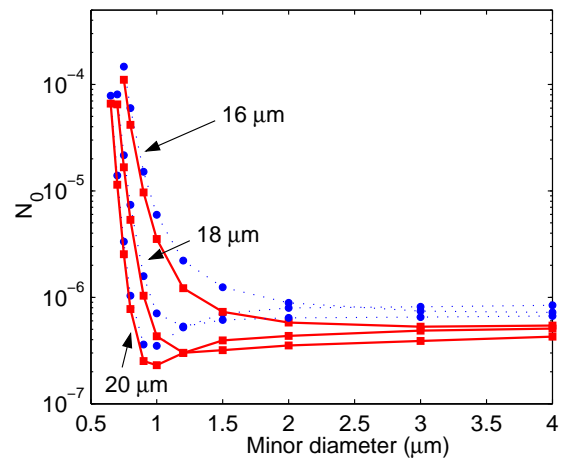}
\end{center}
\caption{Critical atom number $N_{0}$\ vs. minor diameter for a toroidal
microcavity with principal diameters of 16, 18, and 20 microns. For small
minor diameters the critical atom number decreases as the principal diameter
increases. Both TE (blue circles) and TM (red squares) polarizations are
shown. The plots indicate that there is a minimum value of the critical atom
number near $2\times10^{-7}$ for a toroidal cavity with a principal diameter
of 20 microns and an inner diameter of 1 micron (TM mode).}%
\label{critical_atom}%
\end{figure}

One of the primary reasons high-Q whispering-gallery-mode cavities are
promising for cQED is their very low critical atom number. Figure
\ref{critical_atom} shows the calculated critical atom number versus minor
diameter for toroid principal diameters of 16, 18, and 20 microns. The plot
shows that for the larger principal diameters of 18 and 20 microns there is a
minimum in the critical atom number as the toroidal minor diameter is reduced.
The minimum occurs near a minor diameter of 1 micron. This minimum arises from
the clamping of the total quality factor (to the quality factor resulting from
silica absorption) for larger minor diameters when the principal diameter is
greater than approximately 18 microns. Thus, by reducing the minor diameter
for a fixed principal diameter, the quality factor is nearly unchanged while
the coupling strength is monotonically increasing. The critical atom number
decreases until the region where the minor diameter is such that the overall Q
is determined by whispering-gallery loss. At this point the critical atom
number increases approximately exponentially. The plot for the 20 micron
principal diameter shows that in a toroidal geometry slightly larger principal
diameters can offer some benefit, as the minor diameter can be compressed more
strongly while maintaining high radiative quality factors, and thereby
lowering the critical atom number. A critical atom number of approximately
$2\times10^{-7}$ is possible using a toroid principal diameter of 20 microns
and a minor diameter of 1 micron.

\section{Experimental measurement of microtoroids for strong-coupling cavity
QED at 852 nm}

The presented numerical results indicate that toroidal cavities can
theoretically obtain high values of atom-cavity coupling while simultaneously
retaining an extremely low critical photon number and in particular an
exceedingly small critical atom number. While in principle the critical atom
number can be more than 100 times smaller than any currently demonstrated
cavity, the necessity of realizing material-limited quality factors exceeding
20 billion is experimentally challenging. The current record for any cavity is
9 billion \cite{Vernooy}, in a large diameter microsphere cavity, whereas for
toroidal cavities quality factors as high as 400 million at a resonance
wavelength of 1550 nm have been realized \cite{UHQ_toroid_APL}. However, for
cavity quality factors much larger than 100 million, the dominant dissipative
mechanism in the atom-cavity system is the radiative decay rate of the atomic
medium, which is 2.61 MHz for the $D_{2}$ transition of cesium. For this
reason more "modest" quality factors, in the range of current experimentally
achievable values (e.g. a few hundred million) are attractive. As these values
are currently realizable for toroidal cavities at a wavelength of 1550 nm, we
have investigated experimentally the quality factors and fabrication limits
for structures designed for strong-coupling to the cesium transition at a
wavelength of 852 nm.

As toroidal cavities are fabricated using a combination of lithography and a
silica reflow process, the advantages of lithographic control and parallelism
are obtained, and in fact are a significant step forward over spherical
cavities. As the shape of the initial silica preform dictates the maximum
possible principal and minor diameter, and is lithographically formed, precise
control of the geometry dimensions is possible. Reproducible principal
diameters ranging from $>$100 microns to 12 microns have been fabricated. This
lower value, while currently dictated by the available laser power in our
setup, is sufficient to obtain the range of principal diameters optimally
suited for cQED, as indicated above. While the capability to obtain
reproducible principal diameters is a significant improvement over spherical
cavities, the ability to accurately control the minor diameter is particularly
important to cQED. As noted previously \cite{Toroid}, the final minor diameter
of the fabricated structures is a result of a combination of factors, which
are the initial silica preform thickness, the supporting pillar size, and the
laser irradiation intensity and duration. Minor diameters as small as 3
microns at principal diameters as low as 12 microns have been realized experimentally.

\begin{figure}[ptb]
\begin{center}
\includegraphics[scale=0.85]{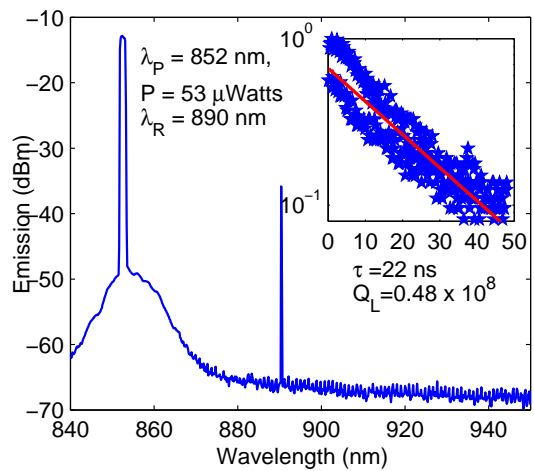}
\end{center}
\caption{Experimental measurement of the intrinsic quality factor for a
toroidal microresonator at a resonance wavelength of 852 nm. The main figure
shows the generation of stimulated Raman scattering, illustrated by the
secondary Raman peak located at a wavelength of 890 nm. The threshold pump
power for stimulated Raman scattering (53 microwatts) can be used to infer the
intrinsic quality factor of $1\times10^{8}$ for this cavity. The inset shows
the temporal cavity decay resulting from a series of ringdown measurements for
a different toroidal microcavity. The measured photon lifetime of $\tau=22$ ns
corresponds to a loaded quality factor of $Q_{L}=0.48\times10^{8}$. After
correcting for fiber-taper loading and the presence of backscattering, an
intrinsic quality factor of $1.2\times10^{8}$ is obtained.}%
\label{experiment}%
\end{figure}

We have measured the quality factor of a series of fiber-taper-coupled
toroidal microcavities at a wavelength of 852 nm, using an experimental
apparatus similar to previous work \cite{Spillane_Nature,Toroid}. The
excitation laser was a New Focus Vortex laser with a tunability of 40 GHz with
a center wavelength of 852.359 nm. The laser output was double-passed through
an acousto-optic modulator for the purpose of performing a cavity ringdown
measurement. The resulting beam was able to be extinguished by a
TTL\ electrical control signal, with a corresponding optical decay time of 15
ns. This beam was then coupled into a single-mode 850 nm fiber and
subsequently interacted with the toroidal resonators through the tapered
portion of the fiber. Due to the limited tuning range of the excitation laser
(which is less than the free-spectral range between fundamental modes in the
cavity principal diameters of interest), overlap of a fundamental resonance
with the laser wavelength range was difficult. Obtaining an optical
fundamental mode at 852.359 nm was achieved by thermally-shifting the optical
resonance through the use of a Peltier heating element, which allowed tuning
of the cavity resonance by up to approximately 50 GHz. Upon realization of a
fundamental cavity resonance at the proper wavelength, the intrinsic quality
factor was inferred two ways (figure \ref{experiment}): through cavity
ringdown \cite{Toroid}\ and through the threshold for stimulated Raman
scattering \cite{Spillane_Nature}. The result of both measurements were in
agreement and resulted in a measured quality factor as high as $Q_{total}%
=1.2\times10^{8}$ in a cavity with a principal diameter of 50 microns and a
minor diameter of 6 microns. For this cavity geometry, the whispering-gallery loss is negligible ($Q_{rad}\approx 10^{36}$) compared to the intrinsic silica absorption loss, such that the overall theoretical quality factor can be as high as $Q_{total}\approx 2\times 10^{10}$. We expect upon further measurements that this quality factor can be increased to levels comparable to measurements performed at a wavelength of 1550 nm (400 million).

While this cavity geometry is far from the optimal geometry suggested in this
manuscript, this structure was chosen in order to increase the likelihood of
finding a fundamental resonance at 852 nm. Even for this relatively large
structure, cavity QED parameters of $(g/(2\pi),n_{0},N_{0})=(86$
MHz$,4.6\times10^{-4},1.0\times10^{-3})$ are calculated. Comparison of these
values to current FP cavities \cite{Hood00,SingleAtomLaserKimble,FP}%
\textit{\ }indicates that even without additional improvements in fabrication
these results are close in coupling strength and improved with respect to the
critical atom number. Additionally, if we restrict the geometry and overall
quality factor to values which are currently realizable (i.e., a quality
factor of 100 million at a wavelength of 852 nm with a minor diameter of 3.5
microns, which represents a reasonably comfortable margin from the actual
current limits), the optimal principal diameter is 13 microns (this geometry
has a radiative quality factor of 180 million). For these values the TM
polarized optical mode would have cQED parameters of $(g/(2\pi),n_{0}%
,N_{0})=(450$ MHz$,1.7\times10^{-5},4.5\times10^{-5})$, which are far superior
to current FP's.

\section{Comparison of microtoroids with other resonators for cavity QED}

\begin{table*}[ptb]%
\footnotesize
\begin{tabular}
[t]{|l|l|l|l|l|l|}\hline
Resonator system & Coupling coefficient & Critical photon & Critical atom &
Coupling to & Rate of\\
&  & number & number & dissipation ratio & optical information\\
& $g/(2\pi)$ [MHz] & $n_{0}$ & $N_{0}$ & $g/max(\gamma_{\perp},\kappa)$ &
$R\equiv g^{2}/\kappa$ [bits/sec]\\\hline
Fabry-Perot & 110 & $2.8\times10^{-4}$ & $6.1\times10^{-3}$ & 7.8 &
$5.4\times10^{3}$\\
experimental state-of-the-art &  &  &  &  & \\\hline
Fabry-Perot & 770 & $5.7\times10^{-6}$ & $1.9\times10^{-4}$ & 36 &
$1.7\times10^{5}$\\
projected limits &  &  &  &  & \\\hline
Microsphere experimental & 24 & $5.5\times10^{-3}$ & $3.0\times10^{-2}$ &
7.2 & $1.1\times10^{3}$\\
(D=120 microns) &  &  &  &  & \\\hline
Microsphere theory &  &  &  &  & \\
Maximum $g$ (D=7.25 micron) & 750 & $6.1\times10^{-6}$ & $7.3\times10^{-1}$ &
0.01 & $4.5\times10^{1}$\\
Minimum $N_{0}$ (D=18 micron) & 280 & $4.3\times10^{-5}$ & $3.1\times10^{-6}$
& 107 & $1.1\times10^{7}$\\\hline
Photonic bandgap cavity & 17000 & $7.6\times10^{-9}$ & $6.4\times10^{-5}$ &
3.9 & $5.1\times10^{5}$\\\hline
Toroidal microcavity theory &  &  &  &  & \\
Maximum $g$ & $>700$ & $6.0\times10^{-6}$ & $2.0\times10^{-4}$ & 40 &
$1.6\times10^{5}$\\
Minimum $N_{0}$ & 430 & $2.0\times10^{-5}$ & $2.0\times10^{-7}$ & 165 &
$1.6\times10^{8}$\\\hline
\end{tabular}
\caption{Summary of the relevant parameters for cavity QED for a variety of
resonator systems. The table shows both the experimental state-of-the-art
\cite{BuckPRA} and the projected limits for a Fabry-Perot cavity \cite{FP},
plus current experimental results with silica microspheres \cite{Vernooy_CQED}%
. Furthermore, a theoretical comparison between silica microspheres
\cite{BuckPRA}, photonic bandgap cavities \cite{PBGPainterMabuchi}, and
toroidal microresonators (this work) is also given. The results indicate that
toroidal cavities can uniformly exceed the performance on these parameters for
both FP cavities and silica microspheres. Comparison with PBG cavities
indicates that toroids possess much lower atom-cavity coupling strengths (as a
result of their much larger mode volumes), but still result in greatly
improved critical atom numbers due to their very large quality factors.}%
\label{cavity_comparison}%
\end{table*}

Table \ref{cavity_comparison} presents a comparison of cQED parameters for
various cavity types including toroidal, FP, and photonic crystal. To date,
most experimental work has involved the use of Fabry-Perot cavities, with
current state-of-the-art fabrication technology allowing the attainment of
coupling strengths of 110 MHz, with corresponding critical atom numbers of
$6\times10^{-3}$ \cite{Hood00}. Estimates on the theoretical performance
limits of FP cavities have also been investigated \cite{FP}, predicting
coupling rates as large as 770 MHz, with a corresponding critical atom number
of $2\times10^{-4}$. While this level of performance may be theoretically
possible, the current necessity of expensive and sophisticated high-reflection
dielectric mirror coatings does not bode well for easy improvements with
respect to current technology. This is one of the reasons silica microspheres
are of such high interest. Calculation of the limits possible with silica
microspheres \cite{BuckPRA} shows that not only is it possible to obtain high
values of atom-cavity coupling solely by changing the cavity diameter, which
is easily in the realm of current fabrication capability, but their ultra-high
quality factors result in significant improvement in the critical atom number
(with values approaching $3\times10^{-6}$ possible provided that silica
absorption-limited quality factors can be obtained). Even using quality
factors in the range of a few hundred million, which is already experimentally
demonstrated, critical atom numbers around $10^{-4}$ are possible, which is
comparable to the FP limit. From the analysis of the previous section, we see
that toroidal cavities can attain coupling strengths comparable to or
exceeding the best values possible for either FP or microsphere cavities,
while at the same time providing much lower critical atom numbers. As
discussed previously, this arises from the extra level of geometrical control
possible in a toroidally-shaped cavity, which allows one to retain both the
high-coupling strength representative of small-mode volume cavities while
preserving high quality factors. Clearly this fact, along with other
advantages in control and reproducibility over spherical cavities, suggests
these structures promising for cQED experiments.

Lastly, a comparison with photonic bandgap cavities is also provided in the
table. Due to the realization of optical mode volumes near the fundamental
limit in a dielectric cavity \cite{PBGJelenaCQED}, combined with recent
results demonstrating reasonably high quality factors ($\sim$45000)
\cite{Noda}, these cavities are strong candidates for chip-based
strong-coupling cQED \cite{PBGPainterMabuchi}. While these structures can
potentially achieve atom-cavity coupling strengths $g\gtrsim$17 GHz
\cite{PBGPainterMabuchi}, far greater than those possible in a silica
dielectric cavity, their much lower quality factors results in greater
critical atom numbers than possible in toroidal microcavities. For example,
the work of Ref. \cite{PBGPainterMabuchi} projects $N_{0}=6.4\times10^{-5}$.
We also note that the correspondingly lower quality factors also result in
modest ratios of coupling to dissipation $g/max(\gamma_{\perp},\kappa)$ (a
figure of merit indicative of the number of Rabi oscillations which occur) of
$\sim$4 \cite{PBGPainterMabuchi}, much lower than predicted for toroidal
structures ($\sim$165). Furthermore, we can consider an additional figure of
merit, namely the \textquotedblleft rate of optical information per
atom\textquotedblright\ \cite{Kimble-cQED}, given by $R\equiv g^{2}/\kappa$.
The table indicates that toroidal cavities compare favorably with PBG cavities
in this figure of merit as well.

\section{Conclusion}

Our work has demonstrated that toroidal resonators are promising cavities for
investigation of the coupling of an atomic system to the electromagnetic field
in the regime of strong-coupling. Not only are these structures arguably
simpler to manufacture and control than other structures such as microspheres
and FPs, but also allow integration on a silicon chip, paving the way for the
addition of atom traps \cite{ChipBEC} and waveguides which can enhance the
capability and possibly reduce the experimental complexity of cQED studies.
Furthermore, we note that in addition to the enhanced performance benefit of
having a toroidal geometry, the capability to retain a relatively large
resonator diameter over other structures results in a smaller free-spectral
range (FSR). This allows not only easier tuning of the cavity resonance
location to correspond precisely to the atomic transition wavelength, but also
may allow integration of a supplemental far-off-resonance trap by exciting the
cavity at a multiple of the free-spectral range. The realization of a cavity
with a smaller FSR may allow a closer matching of a secondary resonance
location to the pump wavelength which corresponds to state-insensitive
trapping of atomic cesium \cite{PRLTrapping}, which can simplify the
atom-cavity dynamics. The use of a silica dielectric whispering-gallery-cavity
also allows operation over a broad range of wavelengths, with very high
quality factors possible for nearly all resonances. This is in strong contrast
to the mirror reflectivity limits of coated FP cavities.

The ability to connect distant quantum nodes with high efficiency, preferably
over optical fiber, is very desirable for quantum networks. Using FP's,
optical fiber coupling is possible, however the overall coupling efficiency is
modest ($\sim$70\%). Fiber-taper-coupled microtoroids allow coupling
efficiencies in excess of 99\% \cite{Ideality}, above both FP's and PBG
cavities (97\%) \cite{Painter_PC_coupling_efficiency}. This capability to
obtain near complete input and output coupling efficiencies strongly suggests
the use of fiber-coupled silica whispering-gallery-cavities, such as
microtoroids, as building blocks to enable high-performance quantum networks.

As a further note, the use of higher-index contrast dielectric material, can
allow additional improvements in the performance of these structures. The use
of silica as the dielectric of choice in both the spherical geometry and in
the toroidal microcavities studied in this work was convenient, as these
structures not only possess record high quality factors but are currently
producible. However, as the radiative quality factor of a
whispering-gallery-type cavity is strongly dependent on the refractive index
difference between the structure and the external environment, much smaller
modal volumes are possible for a given quality factor with the use of a
higher-index resonator material. In fact, this is one of the reasons PBG
cavities fabricated from silicon or other high-index dielectrics can obtain
ultra-small mode volumes. A simple comparison of the mode volume possible in a
silicon toroid shows that a mode volume on the order of only about 10 times
larger than PBG cavities is possible, with much higher quality factors. While
this work has focused on silica microcavities, the reflow process is a
relatively flexible method, thus suggesting that it may be possible to also
create high-index ultra-high-Q quality factor cavities which come closer to
the large coupling strengths of PBG cavities while further improving the
critical atom number.

Lastly, the current experimental ability to obtain large coupling strengths
with quality factors exceeding 100 million is promising for the immediate use
of these structures in strong-coupling studies. We are currently pressing
forward on improving the fabrication capabilities and losses of these
structures. Coupled with the intrinsic fiber-optic compatibility of these
structures, and the demonstration of near lossless excitation and extraction
of optical energy from these structures using tapered optical fibers
\cite{Ideality}, toroidal microcavities can provide a highly advantageous
experimental system for the investigation of strong-coupling cavity QED.

\section{acknowledgements}
KJV -- This work was supported by DARPA, the Caltech Lee Center,
and the National Science Foundation. HJK -- This work was supported by the National Science Foundation, by the
Caltech MURI Center for Quantum Networks, by the Advanced Research and
Development Agency, and by the California Institute of Technology.

\end{document}